\documentclass{acm_proc_article-sp}
\pdfoutput=1

\usepackage{amssymb}
\usepackage{graphicx}
\usepackage{subfigure}
\usepackage{color}
\usepackage{listings}

\usepackage{multirow}

\usepackage{url}

{\end{list}}

{\end{list}}

{\end{list}}

{\end{list}}

{\end{list}}

{\end{list}}

{\end{list}}

{\end{list}}

{\end{list}}

{\end{list}}

{\end{list}}

{\end{list}}

\begin{document}

\title{Component Based Clustering in Wireless Sensor Networks}



\author{
%
%
\alignauthor
Dimitrios Amaxilatis $\ \ $ Ioannis Chatzigiannakis $\ \ $ Christos Koninis $\ \ $ Apostolos Pyrgelis \\
       \affaddr{Research Academic Computer Technology Institute}\\
       \affaddr{and Computer Engineering and Informatics Department, University of Patras}\\
       \affaddr{Greece}\\
       \email{\{amaxilat,ichatz,koninis,pyrgelis\}@cti.gr}
}


\maketitle
%
%

\maketitle

\begin{abstract}
Clustering is an important research topic for wireless sensor networks (WSNs). A large variety of approaches has been presented focusing on different performance metrics. Even though all of them have many practical applications, an extremely limited number of software implementations is available to the research community. Furthermore, these very few techniques are implemented for specific WSN systems or are integrated in complex applications. Thus it is very difficult to comparatively study their performance and almost impossible to reuse them in future applications under a different scope. In this work we study a large body of well established algorithms. We identify their main building blocks and propose a component-based architecture for developing clustering algorithms that (a) promotes exchangeability of algorithms thus enabling the fast prototyping of new approaches, (b) allows cross-layer implementations to realize complex applications, (c) offers a common platform to comparatively study the performance of different approaches, (d) is hardware and OS independent. We implement $5$ well known algorithms and discuss how to implement $11$ more. We conduct an extended simulation study to demonstrate the faithfulness of our implementations when compared to the original implementations. Our simulations are at very large scale thus also demonstrating the scalability of the original algorithms beyond their original presentations. We also conduct experiments to assess their practicality in real WSNs. We demonstrate how the implemented clustering algorithms can be combined with routing and group key establishment algorithms to construct WSN applications. Our study clearly demonstrates the applicability of our approach and the benefits it offers to both research \& development communities.
\end{abstract}



\section{Introduction}
\label{sec:intro}

During the last decade, \emph{wireless sensor networks} (WSNs) gained the interest of computer science, industries and academia, not only from theoretical but also from practical perspectives~\cite{wsn}. Consisting of spatially distributed autonomous sensor-equipped devices, WSNs allow the cooperative monitoring of physical or environmental conditions (e.g., temperature, light, pollutants, etc.), enabling a multitude of applications in both urban and rural contexts. 

Many of the proposed applications assume large node populations densly deployed over sizeable areas. Thus far, we have only a few examples of large-scale deployments of such systems in rural context, whereas in the case of urban deployments we are currently seeing great advances, as signified by research projects such as CitySense~\cite{CITYSENSE} and SmartSantander~\cite{SMARTSANTANDER}. 

Current wireless communication technologies used by the vast majority of off-the-shelf sensor nodes allow short range message exchanges. Thus the network is operated in multi-hop fashion to enable scalable routing, data aggregation, and querying. Designing and operating such large size networks requires scalable architectural and management strategies. Additionally, it is of high importance to  design energy-aware algorithms for preserving the network lifetime.

Since \cite{baker2002architectural}, grouping sensor nodes into clusters has been widely pursued by the research community in order to achieve network scalability, fault-tolerance and energy efficiency. In WSNs a large variety of approaches have been presented focusing on different performance metrics. Some have been proposed as stand alone methods (see  e.g.,~\cite{Iwanicki:EWSN:2009}), others incorporated  as sub protocols in larger solutions designed to solve more specific problems such as query execution, aggregation, localization etc. (see e.g.,~\cite{Iwanicki:IPSN:2009,DBLP:journals/tmc/YounisF04}). 

Unfortunately, even though all of them have many practical applications, an extremely limited number of software implementations for real sensor nodes is available to the community  \cite{Younis05anexperimental,Iwanicki:IPSN:2009,Iwanicki:EWSN:2009}. Furthermore, these very few algorithms are implemented for specific WSN systems or are integrated in complex applications,  therefore rendering them very difficult to reuse by future application developers. For example, the authors of \cite{Iwanicki:IPSN:2009} implement the three cluster hierarchy maintenance algorithms presented in \cite{Iwanicki:IPSN:2009,Iwanicki:EWSN:2009} as part of a routing module for TinyOS v2.0. Similarly, the work in \cite{Younis05anexperimental} implements the algorithm of \cite{DBLP:journals/tmc/YounisF04} by extending the multi-hop routing module provided in TinyOS v1.0 and by modifying the well-known Surge application in order to take advantage of the clustering scheme. Since these implementations are strongly coupled with the routing modules, using them for a different purpose (e.g., data aggregation, energy efficiency, localization etc.) requires significant development effort. 

Surprisingly, to the best of our knowledge, these are the only clustering schemes that are implemented in a contemporary OS for WSNs. All other implementations of clustering schemes (e.g., the well known LEACH protocol \cite{HCB00}) are in simulated environments (e.g., ns2) and the vast majority are not publicly available and/or open source. This inevitably makes it very difficult to comparatively study the performance of existing approaches so that to select the best approach for a particular application scenario. Moreover, even converting these implementations into code for real devices (e.g., with 8-bit microprocessors and tiny amounts of RAM) is certainly a non trivial task.

In this work, we start by examining a $90$ recent clustering algorithms for WSNs in order to identify the key components that re-appear in the majority of the cases. We carefully identify two basic procedures (that of \textit{Cluster-head selection} and \textit{Node grouping}) that appear in almost every algorithm of the $90$ studied. We then propose a component-based architecture for developing clustering algorithms. We focus on $26$ well studied clustering algorithms and design a very limited set of base modules that is highly parameterizable and can be interchanged to implement all these algorithms or new ones. Essentially, we totally avoid implementing each clustering algorithm as a monolithic, stand-alone piece of code. The component-based approach allows us to promote \textbf{exchangeability} of different modules that interact using well-defined interfaces. Modules can be exchanged with other implementations without affecting the remaining code. 

Out of the $90$ algorithms, we select the $5$ most characteristic and implement them under our architecture. These algorithms are carefully selected so that each of them contains as many unique design choices as possible. We conduct an extended simulation study to demonstrate the faithfulness of our implementations when compared to the original implementations. For all cases, the resulting implementations achieve almost identical performance to the one acquired by the original studies. We exploit the benefits of our architecture and also conduct experiments in a hardware testbed in order to assess their performance in real conditions. The implementation effort required to develop these algorithms under our modular approach is significantly short; e.g., for the case of the well known LEACH~\cite{HCB00} algorithm, only $2\%$ of the final lines of code is unique, the rest are reused from other, common modules.

Another benefit of our approach is that we provide a common platform so that comparisons between algorithms is easily accessible. Developers can easily mix and match modules in order to provide new clustering variants that best fit their application specifications. The ability to implement new algorithm variants with minimum effort is of significant importance for conducting experimental-driven research. 

Our approach also offers the ability to implement \textbf{cross-layer algorithms}. The modular design proposed allows to use the clustering algorithms as sub-protocols in other problems such as energy conservation, routing, role assignment, security etc.. This reduces the application development effort and also simplifies the implementation of more complex schemes.

All implementations are done using \textsf{Wiselib} \cite{DBLP:conf/ewsn/BaumgartnerCFKKP10}: a code library, that allows implementations to be OS-independent.  It is implemented based on C++ and templates, but without virtual inheritance and exceptions. All implemented algorithms  are \textbf{platform independent} as they can be compiled on a number of different hardware platforms (e.g., TelosB, iSense, ScatterWeb) and \textbf{OS independent} as they can be automatically used in systems implemented using C (Contiki), C++ (iSense), and nesC (TinyOS). 

Finally, we conduct a thorough evaluation using both simulated and experimental environments. For all cases, our simulations are at very large scale thus also demonstrating the scalability of the original algorithms beyond their original presentations. The results of the evaluation also indicate that our implemented code achieves high \textbf{scalability} and \textbf{efficiency}. Moreover, for all cases, for the first time, we conduct experiments and assess their practicality in real WSN. This also demonstrates the capability of our approach to make code easily available to the community that are hardware and OS independent.

\section{Clustering Techniques}
\label{sec:clustering}

A large variety of clustering algorithms has been presented during the past years. In the relevant bibliography there exist several surveys and tutorials (e.g. 
\cite{AY07, MamalisChapter, 1688699, CLL}) that attempt to categorize and classify the various protocols based on the design choices and mode of operation.

For example, in \cite{AY07} they survey $54$ algorithms and identify the relevant architectural parameters that play an important role in their design, such as the \textit{Network dynamics}, \textit{In-network data processing} and \textit{Node deployment and capabilities}. They classify the protocols based on their main objectives in the following categories: \textit{Load balancing}, \textit{Fault-tolerance}, \textit{Increased connectivity and reduced delay}, \textit{Minimal cluster count}, \textit{Maximal network longevity}. Finally, they provide a taxonomy of the examined algorithms that takes into account the \textit{Cluster properties}, \textit{Cluster-head capabilities} and \textit{Clustering process}. Similarly, in \cite{MamalisChapter} they refer to $44$ clustering algorithms. They mention the most important parameters for the clustering procedure which are \textit{Cluster Count}, \textit{Cluster Overlap}, \textit{Cluster-head Selection}, \textit{Node Mobility} and \textit{Time Complexity}. Then, they distinguish the clustering algorithms in two main categories: \textit{Probabilistic} and \textit{Non-probabilistic}, depending on the cluster formation criteria and parameters used for cluster-head selection.

By carefully studying $8$ surveys and tutorials (\cite{AY07, MamalisChapter, 1688699, CLL, DecheneSurvey, DBLP:conf/ccece/CN06, 1553578, Erciyes_graphtheoretic}) that present a total of $90$ clustering algorithms, we identify the procedures of \textit{Cluster-head selection} and \textit{Methodology of node grouping} as part of almost every algorithm.  This observation was of great value for our generic algorithm engineering work. Of course, the implementation of these procedures depends on the clustering algorithm itself (e.g. probabilistic/deterministic). In the next subsections, we concentrate on well established clustering algorithms that are examined in the surveys and we present the techniques they employ for the above procedures.

\medskip

\subsection{Cluster-head Selection}

On Table \ref{tab:org}, one can see the techniques used by popular clustering algorithms for the process of cluster-head selection. In some algorithms, like \cite{HCB00, DBLP:conf/infocom/BandyopadhyayC03} each node is assigned a probability $p$ of becoming a cluster-head. In algorithms like \cite{123456, DBLP:conf/icdcs/BakerE81, DBLP:journals/tmc/YounisF04}, some deterministic criteria like node connectivity, node identity and energy are respectively used for electing cluster-heads. Finally, algorithms like \cite{DBLP:journals/cluster/ChatterjeeDT02, DBLP:conf/dcoss/DingHC05}  are based on the combination of criteria in order to assign weights to nodes and decide the cluster-heads based on those.

\begin{table}[h]
\begin{center}
\footnotesize
\begin{tabular}{| p{3.7cm} | p{4.1cm} |}
\hline
Implementation Method & Related Work \\ \hline
  Random-Probabilistic  &\cite{HCB00}, \cite{DBLP:conf/globecom/YoussefYYA06}, \cite{DBLP:conf/infocom/BandyopadhyayC03}, \cite{DBLP:conf/ipps/ManjeshwarA01}, \cite{DBLP:journals/tmc/YounisF04},
 \cite{DBLP:journals/ijdsn/SelvakennedyS07}, \cite{DBLP:journals/ijsnet/LiZ07} \\ 
 Node Connectivity & \cite{123456}, \cite{1115494}, \cite{DBLP:conf/ewsn/ChanP04}\\
 Identity-Based & \cite{DBLP:conf/icdcs/BakerE81}, \cite{DBLP:conf/infocom/AmisPHV00}, \cite{DBLP:conf/infocom/BanerjeeK01} \\
 Energy & \cite{DBLP:journals/tmc/YounisF04}, \cite{DBLP:journals/ijdsn/SelvakennedyS07}, \cite{DBLP:journals/ahswn/YeLCW07}, \cite{1210619}, \cite{DBLP:conf/icpads/XuQ04},  \cite{DBLP:journals/winet/GuoL09}\\
 Sensing Attribute & \cite{DBLP:journals/tosn/YoonS07} \\ 
 Weight-Based & \cite{DBLP:journals/cluster/ChatterjeeDT02}, \cite{DBLP:conf/dcoss/DingHC05}, \cite{Virrankoski05tasc:topology}\\
\hline	
\end{tabular}
\caption{\label{tab:org} Cluster-head Selection Techniques of Well Established Clustering Algorithms}
\end{center}
\end{table}

If someone wishes to abstract the above techniques and group them based on their high-level behavior, two categories may be formed: a probabilistic one and an attribute based one. The probabilistic approach selects cluster-heads based on some probability $p$ whereas the attribute based approach accepts a parameter as input value (e.g., identity, energy, weight) and discovers 
the appropriate cluster-heads.

\subsection{Node Grouping}

When a node is elected as cluster-head, it advertises itself to neighboring nodes in order for them to join its cluster. A node can decide to join a cluster based on various criteria which can be seen on Table \ref{tab:org1}. In most algorithms, e.g., \cite{HCB00, DBLP:conf/globecom/YoussefYYA06, DBLP:conf/infocom/BanerjeeK01}, the criterion for a node  to join a cluster is the distance to the cluster-head. In other algorithms, like \cite{DBLP:journals/ahswn/YeLCW07, DBLP:journals/ijdsn/SelvakennedyS07,  DBLP:journals/cluster/ChatterjeeDT02} the nodes decide to join a cluster-head based on some cluster-head attribute like remaining energy, time to live, etc.. 
 
\begin{table}[h]
\begin{center}
\footnotesize
\begin{tabular}{| p{3.7cm} | p{4.1cm} |}
\hline
Implementation Method & Related Work \\ \hline
 Distance to CH & \cite{HCB00}, \cite{DBLP:conf/globecom/YoussefYYA06}, \cite{DBLP:conf/infocom/BanerjeeK01}, \cite{1115494}, \cite{DBLP:conf/icdcs/BakerE81}, \cite{DBLP:conf/infocom/BandyopadhyayC03}, 
 \cite{DBLP:journals/tmc/YounisF04}, \cite{DBLP:conf/infocom/BanerjeeK01}, \cite{123456}, \cite{DBLP:conf/infocom/AmisPHV00}, \cite{DBLP:conf/ipps/ManjeshwarA01} \\
 CH's Energy &  \cite{DBLP:journals/ahswn/YeLCW07}, \cite{1210619}, \cite{DBLP:conf/icpads/XuQ04}, \cite{DBLP:journals/winet/GuoL09}\\
 SNR to CH & \cite{DBLP:journals/ijsnet/LiZ07} \\
 CH's Time to Live & \cite{DBLP:journals/ijdsn/SelvakennedyS07} \\
 CH's Sensing Attribute & \cite{DBLP:journals/tosn/YoonS07}\\
 CH's Weight & \cite{DBLP:journals/cluster/ChatterjeeDT02}, \cite{DBLP:conf/dcoss/DingHC05}, \cite{Virrankoski05tasc:topology} \\
\hline
\end{tabular}
\caption{\label{tab:org1} Node Grouping Techniques of Well Established Clustering Algorithms}
\end{center}
\end{table}

As can be seen on Table \ref{tab:org1}, various criteria can be used for joining a cluster. Thus, most algorithms require some kind of algorithmic method in order to spread the necessary information into the network. Considering this fact, when trying to abstract the functionalities of existing algorithms regarding node grouping two approaches arise depending on the way the information is propagated in the network: the breadth first discovery (BFS) and the depth first discovery (DFS). For example, in algorithms where the distance to CH is the criterion for grouping, the BFS approach is suitable whereas in algorithms that employ a criterion based on some joint computation (e.g. cluster's average energy or cluster's total weight), a sequential traversal of every node is required and thus the DFS approach is more appropriate. Moreover, the execution of each of these distributed traversal approaches is controlled by the desired hop-distance from the cluster-head.


\section{Architecture}

The analysis of the related previous work reveals the vast number of clustering algorithms that have been proposed. It also reveals the extremely limited number of implementations that exist. These implementations follow a \textit{monolithic} approach: they are implemented as a single software module. In fact, all implementations integrate the resulting single piece of code with the routing protocol of the operating system. Such a heavy coupling of the two code modules achieves good results in terms of efficient, hierarchical routing. On the other hand, if someone wants to reuse the clustering code for a different purpose (e.g., data aggregation, energy saving, role assignment, security etc.) a large part of the code must be modified. Furthermore, if someone wishes to use a different clustering algorithm for routing, again he has to re-integrate it with the existing code.  

Another aspect of algorithm implementations for WSN is the fact that currently there are different hardware and software platforms available. The research community has not converged into a single software or hardware architecture. Therefore, it is of great importance to make sure that code can execute in different software and hardware platforms. Clearly, the monolithic implementations target specific WSN architectures. To make them available for a different platform we have to re-implement the code.

Based on the above it is evident that we have to provide abstractions at two
different levels: (i) at protocol level so that \textbf{exchangeability} of
algorithms and \textbf{cross-layer implementations} can be achieved and (ii) at
architecture level so that \textbf{platform independence} and \textbf{OS
independence} are achieved. 

\subsection{Wiselib: A Generic Algorithm Library for Heterogeneous Sensor Networks}

Our goal is to tackle these problems by following a component-based approach. A central decision to achieve this is to use the \textsf{Wiselib} \cite{DBLP:conf/ewsn/BaumgartnerCFKKP10} algorithm library. It is implemented based on C++ and templates, thus our code can be generic and parameterizable. The selection of particular target OS and hardware platform is done at compile time automatically and efficiently by the library. 

The \textsf{Wiselib} is an algorithm library for sensor networks that is completely written in C++, and uses templates in the same way as Boost and CGAL. \textsf{Wiselib} provides a generic interface to the OS, which simplifies the development process and decreases the need for dealing with low-level functionality of specific hardware platforms. This makes it possible to write generic and platform independent code that is very efficiently compiled for various platforms, such as iSense or Contiki, or the sensor network simulator Shawn \cite{KPBFF05}. As an example we refer to the {\sc send} routine, which while exists in every sensor, it is implemented in a different way for almost every unique hardware. By providing \textsf{Wiselib} with the {\sc Radio} aspect of the hardware which contains the send routine, we can use it in our code. Thus, by making the {\sc Radio} a template type with templated routines,  we can pass different implementations to the library. Essentially, this allows us to truly implement our algorithms once and be able to execute them in most popular hardware platforms with just a simple recompilation.

\subsection{Basic Components}

\begin{figure}
\includegraphics[width=\columnwidth]{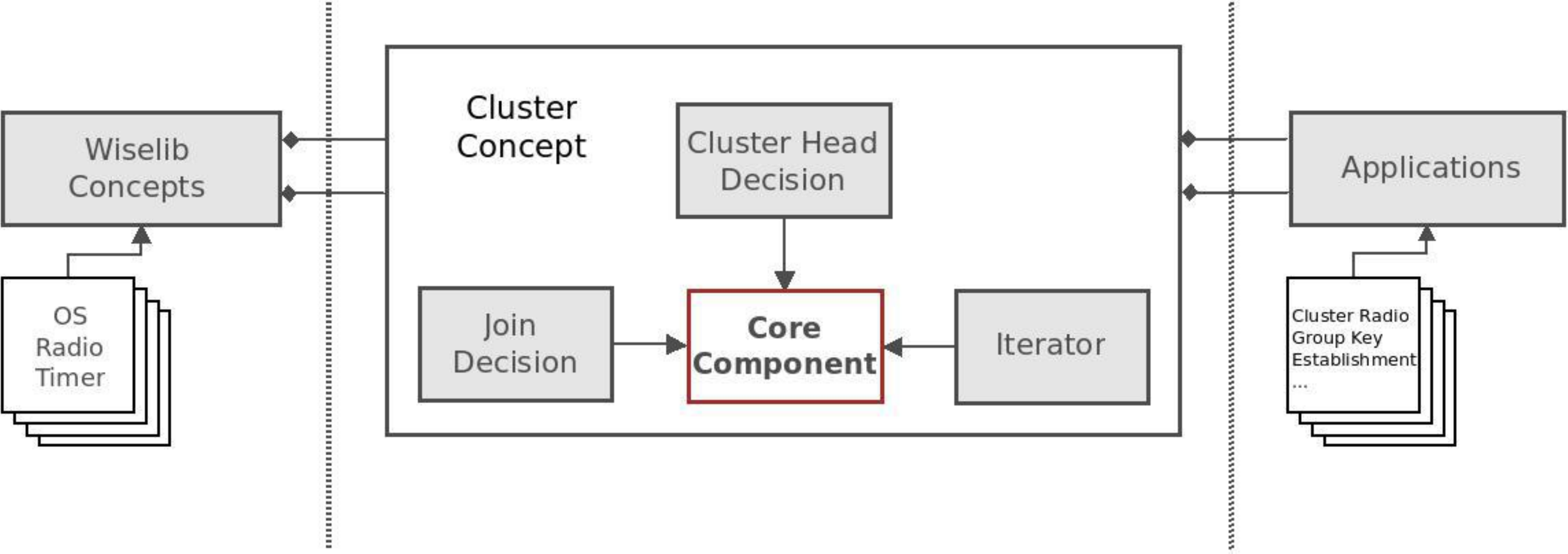}
\caption{\label{fig:arch} Basic components and relation with \textsf{Wiselib}}
\end{figure}

The component-based design that we propose is depicted in Fig.~\ref{fig:arch}.
We partition the logic of clustering algorithms into three pieces with clear
boundaries in terms of functionality provided. Each partition is designed so
that it can progress its work in a relatively independent manner while ensuring
that the correct functionality of the algorithm. Clean interfaces are
provided so that the partitions can easily communicate, fast and without heavy
information exchange.

\noindent\textbf{Cluster-head Decision (\textsc{CHD}).} The first partition that
we propose is related to the cluster-head selection process. All clustering
algorithms have a mechanism for selecting cluster-heads. We wish to implement
such mechanisms as a single, stand-alone, software component. Each particular
implementation takes into account the specific design choices of the clustering
algorithm to select which nodes will become cluster-heads. If the algorithm
proposes periodic rotations of cluster-heads, the component can use
\textit{call-backs} to the \textsc{Timer} component so that it is re-executed
periodically. If the algorithm is cluster-headless, the resulting implementation
will just be an ``empty'' implementation.


\noindent\textbf{Join Decision (\textsc{JD}).} The second partition is related to the methodology by which nodes decide to join cluster-heads. This component constructs the necessary payloads for the \textsc{JoinRequest/JoinDeny/JoinAccept} messages, also it determines if a node will join a cluster when a \textsc{JoinRequest} message is received. The decision can be based on local criteria (e.g., energy levels, mobility status, etc.) and/or it can be related to information provided by the \textsc{JoinRequest} message (e.g., size of cluster etc.). In some other cases the decision may be already taken by the \textsc{CHD} component. 

\noindent\textbf{Iterator (\textsc{IT}).} The third partition is related to the organization of the nodes while clustering decisions are made by each node. This component is responsible for categorizing and storing neighbors into nodes that have already joined the cluster, nodes that have not joined the cluster yet and nodes that have joined another cluster. Collected information is maintained in \textit{membership tables} by the \textsc{IT} component. These tables are of crucial importance for the algorithms that will be executed on top of the clustering -- they are necessary to ensure cross-layering. This component also monitors the node's neighborhood and updates the \textit{membership tables} based on observed changes to the network. This information can be used from the other components. For example, changes in the neighborhood that indicate that a node has left a cluster can trigger a new join decision process from the \textsc{JD} component.

The above three components are used by the main component which we call the \textbf{Core Component (\textsc{CC})}. It is the kernel of our architecture that controls and coordinates all other components so that clusters are properly formed and maintained. The \textsc{CC} provides a public interface for other algorithms to take advantage of the resulting network organization. This public interface \textit{is an implementation of the \textsf{Wiselib} concept of Clustering} and thus provides the cluster's ID, the ID of the cluster-head, and also allows to register a function callback in order to be able to deliver events to external components whenever an change to the cluster occurs, e.g., when the node joined a new cluster, or a neighbor from different cluster was discovered, or a new cluster was formed etc.. In the following we present the life-cycle of \textsc{CC} when forming a new cluster.

\begin{enumerate}

\item \textsc{CHD} is invoked to determine if the node will become a cluster
head or not.

\item \underline{If the node is a cluster-head}: \textsc{JD} is invoked to send \textsc{JoinRequest} messages to nearby nodes and invite them to the cluster. The \textsc{JoinRequest} messages can be sent to all available nodes using Broadcast messages or to selected nodes using the selected nodes ids.



\item Upon receiving a \textit{Join Request} message, \textsc{CC} isolates the message's payload and passes it to \textsc{JD}. \\
\underline{If \textsc{JD} decides to join}, a \textsc{JoinAccept} message is sent to the originator of the \textsc{JoinRequest} message, \textsc{IT} is notified of the address so for it to be saved as the node's \textit{Cluster-head}. Note that if the protocol dictates that nodes may join a cluster even if they are at multi-hop distance from the cluster-head, then the \textsc{IT} may delay the transmission of the \textsc{JoinAccept} message so that other nodes are examined (e.g., this is done in the case of the depth-first search node traversal).\\ \underline{If \textsc{JD} decides not to join}, a \textsc{JoinDeny} message is generated
along with a payload from \textsc{JD} and passed to the originator of the \textsc{JoinRequest} message.

\item If a \textsc{JoinDeny} message is received, its payload is passed to \textsc{JD} to be examined, in case the neighborhood's conditions are of interest and the \textsc{IT} is notified in order to keep track of which neighbors have joined the cluster and which have not.

\item When all nodes have been examined the membership tables are generated by
the \textsc{IT} and the process of cluster formation completes.

\end{enumerate}

\subsection{Implementation Details}

In the following we present a \textsf{Wiselib} concept for each one of four basic components. The design goal of the concept is to cover many cases while staying generic. This follows from 
the requirement that each module must implement all the methods of a concept.

\noindent\textbf{Core Component (\textsc{CC}) Concept.} The \textsc{CC} concept takes as template parameters a set of components types such as \textsc{Radio}, \textsc{Timer} and \textsc{Debug} that are needed for sending messages, registering events and optionally printing debug messages. The most important parameters are the types for the \textsc{CHD}, \textsc{JD} and \textsc{IT} which the \textsc{CC} will use for the clustering algorithm. The first method that initializes the module also provides instances of the components that the module will use. Then we have two methods for enabling and disabling the module, which is useful when it should only be run in certain points in time. After the module is enabled, the \textsc{find\_head()} method is called and starts the cluster formation. Next, we have a method for setting the parameters of the algorithm, which also sets the parameters for every other component. Then, we have a method for registering a callback in order to get notifications upon events. Finally, \textsc{CC} provides a set of functions to access useful information such as the cluster id, the parent node(if any) etc. The concept is defined in \textsf{Wiselib} as follows:

\begin{lstlisting}[language=C++]
template<typename OsModel,
  typename Radio,
  typename Timer,
  typename Debug,
  typename HeadDecision,
  typename JoinDecision,
  typename Iterator>
class CoreComponent {
public:
  void init(Radio&,Timer&,Debug&,CHD&,JD&,IT&);
  void enable(void);
  void disable(void);

  void set_parameters(parameters_t *);
  void find_head(void);

  template<typename T, void(T::*TMethod)(uint8_t)>
    int reg_changed_callback(T* obj);

  node_id_t parent()
  cluster_id_t cluster_id()
  bool is_cluster_head(void);
  ...
};
\end{lstlisting}

\noindent\textbf{Cluster Head Decision (\textsc{CHD}) Concept.} In the \textsc{CHD} concept we have a method for setting the parameters (e.g., the probability value that the module will use). Additionally, there is the method for calculating if the current node is a cluster-head, and a method to get this result. The concept is defined in \textsf{Wiselib} as follows:

\begin{lstlisting}[language=C++]
template<typename Radio, typename Debug>
class ClusterheadDecision {
public:
  void init(Radio& , Debug& );
  void enable(void);
  void disable(void);

  void set_parameters(parameters_t *);
  bool is_cluster_head(void);
  bool calculate_head();
};
\end{lstlisting}

\noindent\textbf{Join Decision (\textsc{JD}) Concept.} In the \textsc{JD} concept we have a method that gives the hop count from the cluster-head, after the node has joined a cluster. It also provides methods that set the payload for specific types of messages. The minimum requirement is three methods, for the \textsc{JoinRequest}, the \textsc{JoinAccept} and the \textsc{JoinDeny} messages. Finally, we have the method \textsc{join} that is called with a new \textsc{JoinRequest} payload, and it calculates if it is going to join the cluster. The concept is defined in \textsf{Wiselib} as follows:

\begin{lstlisting}[language=C++]
template<typename Radio, typename Debug>
class JoinDecision {
public:
  void init(Radio& , Debug& );
  void enable(void);
  void disable(void);

  int hops();
  void get_join_request_payload(block_data_t *);
  void get_join_accept_payload(block_data_t *);
  void get_join_deny_payload(block_data_t *);
  size_t get_payload_length(int);
  bool join(uint8_t *, uint8_t);
};
\end{lstlisting}

\noindent\textbf{Iterator (\textsc{IT}) Concept.} For \textsc{IT}, we provide methods for getting the cluster id and the parent of the node. Moreover, the \textsc{next\_neighbor()} method allows iterating through the neighborhood of the node. If the neighborhood information is not available, we can register a callback function that the Iterator will call to inform us about changes in the neighborhood. The concept is defined in \textsf{Wiselib} as follows:

\begin{lstlisting}[language=C++]
template<typename OsModel,
  typename Radio,
  typename Timer,
  typename Debug>
class Iterator {
public: ...
  void init(Radio&, Timer&, Debug&);
  void enable(void);
  void disable(void);

  cluster_id_t cluster_id(void);
  node_id_t parent(void);
  node_id_t next_neighbor();

  template<typename T, void (T::*TMethod)(uint8_t)>
    int reg_next_callback(T* obj);

private:
  vector_t cluster_neighbors_;
  vector_t non_cluster_neighbors_;
  node_id_t parent_;
  ...
};    
\end{lstlisting}

\subsection{Base Modules}

In this section, we revisit the clustering techniques as summarized in Sec.~\ref{sec:clustering} in the light of the above components and structure. In the sequel we use the term \textit{component} to refer to one of the basic building blocks of our architecture, as presented above, and the term \textit{module} to refer to an implementation of a particular component (or in \textsf{Wiselib} terminology, a particular \textit{component concept}). We implement a total of six modules with parameterizable functionality in order to cover a wide range of clustering approaches. Therefore, we call them the \textit{base modules} since by combining them we can come up with almost all original clustering algorithms examined. For example, many algorithms propose each node to decide whether to become a cluster-head locally with some probability $p$ (e.g., \cite{HCB00}). Other algorithms propose to do this in a deterministic way using specific attributes (e.g., \cite{DBLP:conf/icdcs/BakerE81} using local ids, \cite{1115494} based on node's connectivity, \cite{DBLP:journals/tmc/YounisF04} using node's remaining energy, or
a weighted combination of such criteria as in \cite{DBLP:journals/cluster/ChatterjeeDT02}). Some propose that this is repeated periodically every time period $t$ (e.g., \cite{HCB00}). Some others propose to select cluster-heads up to $k$-hop distance. Table \ref{fig:base_modules} summarizes the six base modules for each one of the four components. 

We understand that it is clearly impossible to come up with a small set of modules that is generic enough to implement \textit{every possible} clustering algorithm. However, we do propose a small set of modules that can be \textit{easily modified and/or extended} by programmers to speed up the implementation effort of a \textit{large range} of algorithms. Therefore, we attempt to cover as many design flavors as possible while staying generic. In the following section we showcase how we used these modules to \textit{fully implement $5$ well known algorithms} and discuss how \textit{they can be further modified to implement $11$ others}. We believe that this is a solid evidence that our approach can help future developers implement applications that rely on a specific clustering scheme.

\begin{table}[h]
\begin{center}
\begin{tabular}{|l||l|r|c|p{3.1cm}|}
\hline
{\small Comp.}		& {\small Module}	& {\small LOC} 		
& {\small Param.}			& {\small Description}			
\\
\hline
\multirow{5}{*}{CHD}	
  & \textsc{prob}	& 149	& $p$,$t$ & {\small Each node becomes cluster
head with probability $p$ every $t$ seconds} \\ \cline{2-5}
			
  & \textsc{attr} 	& 149	& $v$, $k$, $t$	& {\small The node with the
minimum $V$ value becomes the cluster-head of the $k$ hop neighborhood every $t$
seconds}	\\ \hline
\multirow{4}{*}{JD} 	
  & \textsc{bfs}	& 188	& k	& {\small Forms $k$ hop clusters using
breadth-first network discovery} \\  \cline{2-5}
  & \textsc{dfs}	& 190	& k	& {\small Forms $k$ hop clusters using
depth-first network discovery} \\ \hline
\multirow{2}{*}{IT} 	
  & \textsc{norm}	& 318	&\textendash	& {\small Stores information
about the cluster joined and the network neighborhood} \\ \hline
\multirow{2}{*}{CC} 	
  & \textsc{norm}	& 512	&\textendash	& {\small Controls and
coordinates the operations of the other components} \\
\hline
\end{tabular}
\caption{Implemented Base Modules \label{fig:base_modules}}
\end{center}
\end{table}

\section{Example Implementations}

We now proceed by examining five characteristic algorithms and how they are implemented under our scheme. Interestingly, in all five cases the \textsc{CC\_Norm} module remains unchanged. We note that the partition of the functionality in the four components allows us to find easily the places where we have to implement algorithm specific code. Then, the selection of one of the given base modules helps minimize the implementation. We identify $11$ additional algorithms and explain in details how they can be implemented by pointing the places where the changes need to be made and sketching the actual implementations.  All modules that will be described in the sequel are named under the rule \textsc{name\_role} where \textsc{name} is the name of the algorithm and \textsc{role} can be either \textsc{chd}, \textsc{jd} or \textsc{it}.

\noindent{\textsf{LCA.}} A classic algorithm that forms $k$-hop clusters \cite{DBLP:conf/icdcs/BakerE81}. Every node becomes a cluster-head with a probability $p$. All cluster-heads advertise themselves to every node within $k$ hops and all simple nodes join the cluster of the closest cluster-head. The implementation of \textsf{LCA} is derived without any single change from the base modules \textsc{prob}, \textsc{bfs} and \textsc{norm} of Tab.~\ref{fig:base_modules}. Algorithms like \cite{DBLP:journals/ahswn/YeLCW07,1115494} can also be implemented in a straight forward way. This clearly demonstrates that application designer can come up with prototypes in a very short time frame.

\noindent{\textsf{LEACH.}} A well studied algorithm that periodically rotates cluster-heads \cite{HCB00}. Every node $i$ decides independently on predefined intervals to become a cluster-head with probability ${p_i}$.  The value of ${p_i}$ is based on the node's previous role and the desired number of clusters in the network ($k$). So every node becomes periodically a cluster-head and the cluster-head's extra workload is distributed evenly amongst every node. All cluster-heads advertise themselves as in \textsf{LCA} but only $1$ hop away. Simple nodes decide to join the cluster-head that requires the minimum communication energy (using the RSSI value) and inform the selected cluster-head of their decision. For the implementation of \textsf{LEACH} we modify \textsc{prob} so that $p$ is calculated internally every time the cluster is reformed. Interestingly, for the \textsc{jd} module we use the implementation of \textsf{LCA} (i.e., \textsc{lca\_jd}) with very minor modifications (about $8\%$ of the total lines of code). Finally we use the base \textsc{norm} iterator module as is. Based on these modules we can then implement the algorithm of \cite{DBLP:journals/tmc/YounisF04} and \cite{DBLP:journals/ijsnet/LiZ07} that uses a SNR instead of RSSI for the join condition, while \cite{DBLP:conf/infocom/BandyopadhyayC03} uses the same \textsc{chd} but for $k$-hops.

\noindent{\textsf{TCCA.}} In this algorithm, cluster-heads are selected based on the residual energy \cite{DBLP:journals/ijdsn/SelvakennedyS07}. The cluster-heads are periodically selected based on a fixed probability and by taking into account the available energy of the node. We use the base \textsc{prob} module by modifying it to include this extra design choice. Then the \textsc{jd} module takes into consideration the residual energy of the cluster-head as well as the distance of the node to it. This modification is again very easy to implement by modifying the join criteria. 

\noindent{\textsf{MOCA.}} This algorithm forms $k$-hop \emph{overlapping} clusters \cite{DBLP:conf/globecom/YoussefYYA06}. Cluster-head decision is exactly the same as in the base module \textsc{prob}. For the \textsc{jd} component we use \textsc{bfs} but allow a node to \emph{belongs to all} clusters of up to $k$ hops away in order to achieve overlapping clustering. To complete the implementation we also need  to make changes to the \textsc{norm} iterator component. Due to the fact that more than one clusters are selected the variables and data structures provided by the base module are not appropriate. Instead of using single variables, all normal nodes use a table to store the ids of the cluster-heads they belong and cluster-heads use a table to store the the ids of all adjacent clusters. This algorithm is clearly different from the previous ones. The proposed design guides us to the places where changes need to be made. Once again the base modules are heavily reused. Even for the different iterator component, more than 35\% of the original code is present in this algorithm. Interesting, the code of  \cite{DBLP:conf/ewsn/ChanP04} that also forms overlapping clusters is extremely similar to the code we just discussed.

Many algorithms seem to use ideas similar to those presented above. Such algorithms are presented in \cite{DBLP:journals/winet/GuoL09,DBLP:journals/tosn/YoonS07,DBLP:conf/dcoss/DingHC05,Virrankoski05tasc:topology} that exchange only one attribute over one or more hops at regular intervals or every time a special event is detected. Other algorithms like  \cite{DBLP:journals/cluster/ChatterjeeDT02} exchange multiple attributes like mobility, node degree, time as cluster-head and power consumed. All these algorithms can be incorporated in our library with extremely small development effort.

\noindent{\textsf{MaxMinD.}} In \cite{DBLP:conf/infocom/AmisPHV00} a very different algorithm is presented that forms clusters after examining all nodes at $d-hop$ distance. Cluster-heads are elected after an id exchange phase of $2d$ rounds, using a heuristic function on all received ids. Unlike all previous algorithms the cluster join decision is made automatically and nodes join the cluster of their selected cluster-heads. Every simple node then informs its one hop neighbors of its decision and nodes connected to multiple clusters become gateways. All gateway nodes finally inform their cluster-head of their connections to other clusters during the converge cast phase.

This algorithm represents a totally different cluster formation approach where each node elects its cluster-head and joins the cluster at the same time. This requires very different design on the \textsc{chd} and \textsc{jd} modules. To accommodate the special design of \textsf{MaxMinD} we implement the message exchange using payloads provided by the \textsc{jd} module. The cluster join condition is in fact trivial after the cluster that the node will join is decided. The iterator module then requires some extra payloads for the converge cast phase that \textsf{MaxMinD} employs.

The unique design of \textsf{MaxMinD} gives us a good opportunity to make a first attempt to evaluate the component-based design that we propose. In particular we repeat the implementation by following a stand-alone monolithic approach. So now we can compare the two versions based on their lines of code and the resulting binary size. Indeed, code size is not the best code metric available. Component-based implementations usually require more lines of code since they include class definitions and require additional functions declarations so that information of private members is exchanged between modules. Still, it can give us a rough estimate on the effort needed. So, keeping these points in mind, the monolithic version consists of $1200$ lines of code whereas the component-based version is just over $2000$ lines long. However, in the component-based version almost $1000$ lines were reused from the existing base modules. In terms of code size, the component-based executable is $13738$ bytes long while the monolithic is $11951$. This metric reflects better the growth of the code. It indicates that for the particular algorithm the price (in terms of code size) for carrying out a well organized and modular implementation requires an increase of $14\%$. Clearly, this extra code size is the result of a more readable code which is therefore easier to debug, maintain and extend. We strongly believe that \textsf{MaxMinD} is a characteristic example of how even irregular algorithms (in terms of design) can be incorporated into our component-based design with less effort than implementing it in a stand-alone monolithic way. 

We carry out the same comparison with the other algorithms by also implementing them as monolithic modules. Interestingly, we can see similar difference in code size between each version. In some sense this is expected since all component-based implementations use the same interfaces and thus incorporate similar set of functions with a fixed code size. 

We also note that as the library grows in size, the effort to develop new algorithms will reduce as more modules will become available. However, the real benefit of having a large set of implemented modules is the ability to experimental-drive the design of new clustering algorithms. Specific design choices that work well in particular network types can be easily transfered in new algorithms that also combine other modules with different specifications and goals. Inevitably, the availability of a large set of modules will help us develop sophisticated systems with reduced implementation effort.


\begin{table}[h]
\begin{center}
\begin{tabular}{|l|l||l|r|rr|}
\hline
Algorithm			& Module		& Parent	& LOC	& Diff	& (\%)	\\
\hline
\multirow{3}{*}{\textsf{LCA}}	& \textsc{prob}		& \textsc{prob}	& 149 	& 0	& 0\%		\\
				& \textsc{lca\_jd}	& \textsc{bfs}	& 204	& 16	& 8\%		\\
				& \textsc{norm}		& \textsc{norm}	& 318	& 0	& 0\%		\\
\hline
\multirow{3}{*}{\textsf{LEACH}}	& \textsc{leach\_chd}	& \textsc{prob}	& 159	& 10	& 6\%		\\
				& \textsc{leach\_jd}	& \textsc{lca\_jd}& 204	& 10	& 5\%		\\
				& \textsc{norm}		& \textsc{norm}	& 318	& 0	& 0\%		\\
\hline
\multirow{3}{*}{\textsf{TCCA}}	& \textsc{tcca\_chd}	& \textsc{prob}	& 156	& 12	& 8\%		\\
				& \textsc{tcca\_jd}	& \textsc{bfs}	& 217	& 14	& 7\%		\\
				& \textsc{norm}		& \textsc{norm}	& 318	& 0	& 0\%		\\
\hline
\multirow{3}{*}{\textsf{MOCA}}	& \textsc{prob}		& \textsc{prob}	& 149	& 0	& 0\%		\\
				& \textsc{moca\_jd}	& \textsc{bfs}	& 158	& 69	& 43\%		\\
				& \textsc{moca\_it}	& \textsc{norm}	& 390	& 250	& 64\%		\\
\hline
\multirow{3}{*}{\textsf{MaxMinD}}&{\small \textsc{maxmind\_chd}}& \textsc{attr}& 375	& 320	& 85\%		\\
				&{\small \textsc{maxmind\_jd}}	& \textsc{bfs}	& 237	& 162	& 68\%		\\	
				&{\small \textsc{maxmind\_it}}	& \textsc{norm}	& 528	& 399	& 75\%		\\
\hline 
\end{tabular}
\caption{Implemented Algorithms, inherited Base Modules (see Tab.~\ref{fig:base_modules}) and modified lines of code \label{fig:algorithms}}
\end{center}
\end{table}

\section{Validity of Implementation}
\label{sec:valid}

In this section we wish to evaluate the faithfulness of the implementation of the clustering algorithms under our approach. In order to do this we comparatively study the results of the performance evaluation of the protocols under our approach and the one that their designers did. The comparative study is conducted in a simulated environment created using \textsf{Shawn} \cite{KPBFF05}, a simulation framework that focuses on an abstract, repeatable and expressive approach to WSN simulation. By replacing low-level effects with abstract and exchangeable models, the simulation can be used for huge networks in reasonable time while keeping the focus on the actual research problem. It provides many options such as packet loss, radius of communication, ways of communicating and even mobility in an abstract way, without needing to provide specific code for every hange. 

Although it is very difficult to reproduce the exact topologies for the experiments, we try to generate topologies with as similar characteristics as the ones used in the original publications. Also some minor differences are to be expected since we do not use the same simulation environment. The results of this comparative study is very positive as the values acquired by our performance analysis are very close to the ones acquired from the original studies. Our values seem to confirm a valid behavior of the modular implementations of the protocols. 

\begin{figure}
\includegraphics[width=0.45\textwidth,keepaspectratio=true]{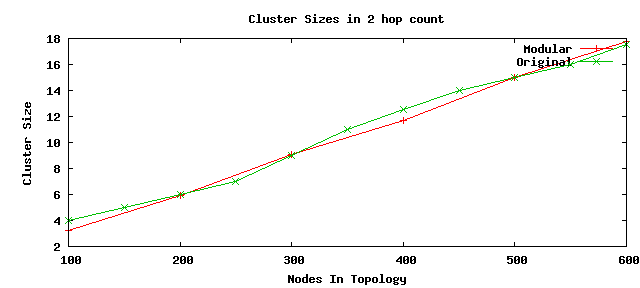}
\includegraphics[width=0.45\textwidth,keepaspectratio=true]{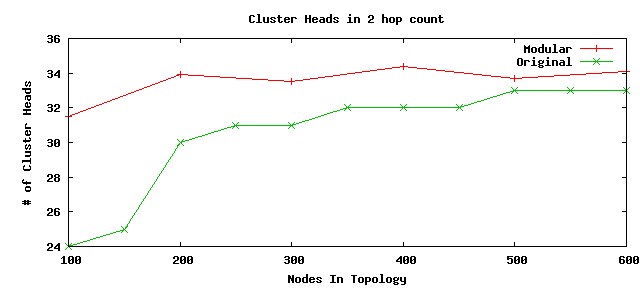}
\caption{Comparison of Original Implementation of \textsf{MaxMinD} and Our Modular Implementation regarding \# of Cluster Heads and Avg. Cluster Size\label{fig:valid:maxmind}}
\end{figure}

For \textsf{MaxMinD}, as we can see in \cite{DBLP:conf/infocom/AmisPHV00}, the original experiments create networks with up to $600$ nodes in a $200 \times 200$ units region and communication range of $20$ units. The parameter $d$ was set to $2$ and the main metrics extracted are the number of cluster-heads generated and the size of the clusters formed. Fig.~\ref{fig:valid:maxmind} depicts the results from our performance analysis (marked as \textit{Modular}) in comparison to the results from the original analysis of conducted in \cite{DBLP:conf/infocom/AmisPHV00}. As we can see from the simulations we performed, the clusters formed have almost identical size to the originals while the cluster-head count is very similar. The minor discrepancy observed in the number of cluster-heads for small topologies is related to the distribution of the node ids in the network. In \textsf{MaxMinD}, the position of each node id is crucial to the number of cluster-heads generated as the heuristic that is used is based on the maximum ids transmitted during each round and sequentially the way the ids are placed. As we do not know the exact position of all node ids used in the original experiments it is difficult to achieve the same result. In our experiments we used a uniform random distribution. 

\begin{figure}
\includegraphics[width=0.45\textwidth,keepaspectratio=true]{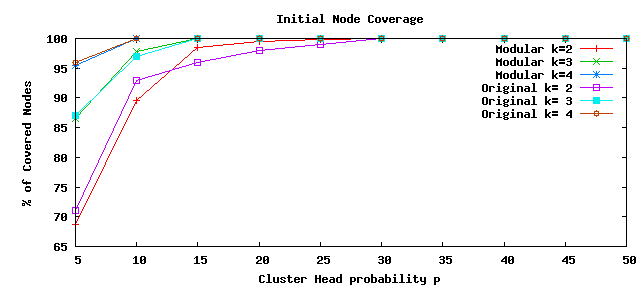}
\includegraphics[width=0.45\textwidth,keepaspectratio=true]{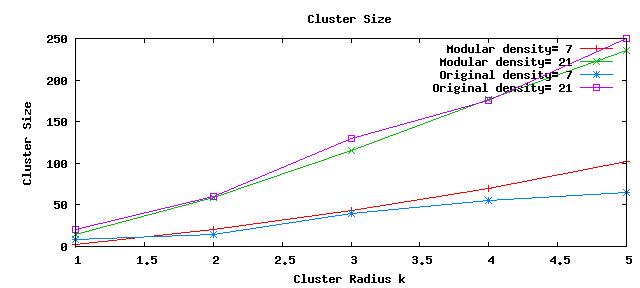}
\caption{Comparison of Original Implementation of \textsf{Moca} and Our Modular Implementation regarding Percentage of Covered Nodes and Avg. Cluster Size\label{fig:valid:moca}}
\end{figure}

For \textsf{Moca}, the simulations performed in \cite{DBLP:conf/globecom/YoussefYYA06} focus on special features of overlapping clustering algorithms. Metrics like the number of covered nodes during the initial phase of the algorithm, cluster overlapping degree and orphan clusters is extracted. The size of the clusters formed is also measured. Again, we try to reproduce the same simulation environment. We create a world with $400$ nodes and a node density of $9$ neighbors. We simulate cluster formation with cluster-head probability of up to $50\%$ and cluster diameters of $2$, $3$ and $4$ hops. As we can see in Fig.~\ref{fig:valid:moca} the percentage of the covered nodes achieved by our modular implementation is almost identical with that achieved by the original implementation. Moreover, in order to evaluate the size of the clusters formed by \textsf{Moca}, we need to generate topologies of $400$ nodes and density of up to $21$ neighbors. We use a fixed cluster-head probability of $15\%$ and increase the diameter of the clusters formed. Once again the results gathered from our evaluation, as shown in Fig.~\ref{fig:valid:moca}, indicate clearly that the performance of our modular implementation is almost identical to that the one achieved by the implementation of the algorithm's designers.

For the other three algorithms, the simulation study conducted in \cite{DBLP:conf/icdcs/BakerE81,HCB00,DBLP:journals/ijdsn/SelvakennedyS07} is very limited and/or is focused on energy related aspects that we do not investigate in this work. Still, even for this limited set of results (e.g., number of elected cluster-heads for a specific topology of $100$ nodes), the corresponding results of the evaluation of our implementations are almost identical. Furthermore, the perforamnce evaluation conducted in the following sections produce results that are justifiable by the qualitative analysis conducted in the original publications of all five algorithms.

\section{Performance Evaluation}
\label{sec:eval}

In this section we evaluate the performance of the base modules based on large-scale simulations so that we can assess their \textit{scalability} and experiments using hardware to assess their \textit{effectiveness} in real environments.

The simulator experiments allow us to evaluate the scalability of the base modules in various network topologies. We work with two different types of network topologies. The first set of topologies have \emph{fixed network diameter}: as we increase the number of nodes the network density increases. The second set of topologies have \emph{fixed node density}: as the number of nodes increases so does the network diameter. In the second set, we generate networks with a density of $8$ (i.e., on average each node is connected with $8$ other nodes) and with density of $15$. The number of nodes for each topology with fixed density varies form $10$ to $10000$ and for the second case with diameters of $1$ to $100$ hops. The fixed diameter topologies start from $10$ nodes with $4$ average neighbors and reach $1000$ nodes with $256$ neighbors and a network diameter of $5$ hops. For higher densities the equivalent graphs become fully connected and thus they are useless for our evaluation. On each experiment we focus on the amount of time each module requires to complete; time its counted in simulation rounds. We also measure the total number of messages sent. Finally, we evaluate the average size of the clusters formed.

\begin{figure}
\begin{center}
\includegraphics[width=0.45\textwidth,keepaspectratio=true]{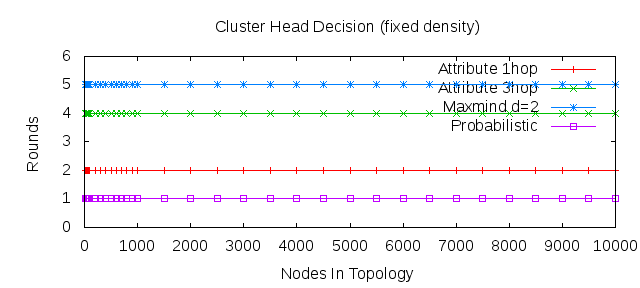}
\includegraphics[width=0.45\textwidth,keepaspectratio=true]{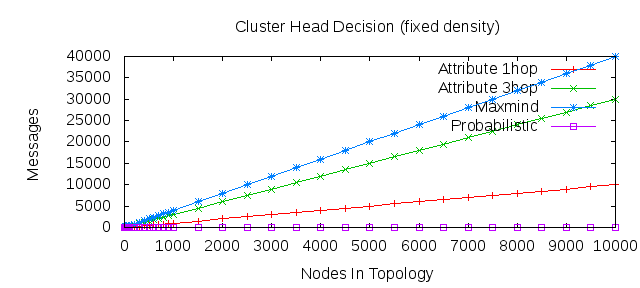}
\caption{Performance of \textsc{chd} modules for different network sizes of fixed density \label{fig:shawn:chd}}
\end{center}
\end{figure}

Through out all the simulations for the cluster-head phase we use the \textsc{attr} to elect cluster-heads with minimum ids in $1$ and $3$ hops, \textsc{prob} and \textsc{maxmind$\_$chd}. For the cluster formation 
we use \textsc{bfs}, \textsc{dfs}, \textsc{maxmind\_jd} and \textsc{moca\_jd}. Finally, we evaluated \textsc{norm}, \textsc{moca\_it} and \textsc{maxmind\_it}. 

We start by assessing the performance of the cluster-head decision modules as depicted in Fig.~\ref{fig:shawn:chd}. Note that similar results hold for the networks of fixed diameter. For space limitations we only include the results from the fixed density networks. As expected, all modules require constant number of rounds to complete. The total time required is in fact determined by the specific parameters. Similarly, the number of messages exchanged is linear to the $k$-hop parameter of the modules. Clearly, the \textsc{PROB} module achieves the best scalability since the decision is local and requires a constant number message exchanges.

\begin{figure}
\begin{center}
\includegraphics[width=0.45\textwidth,keepaspectratio=true]{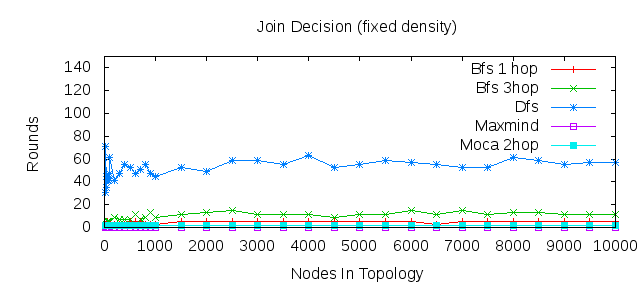}
\includegraphics[width=0.45\textwidth,keepaspectratio=true]{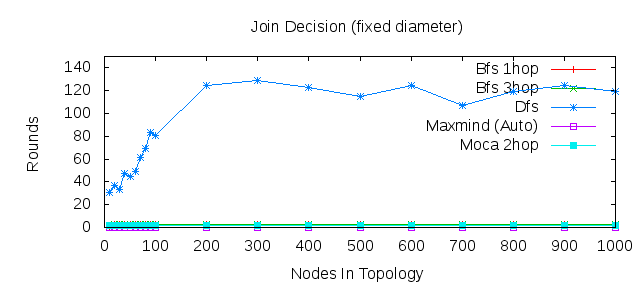}
\includegraphics[width=0.45\textwidth,keepaspectratio=true]{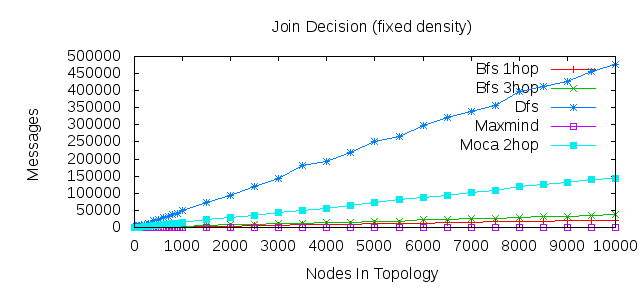}
\caption{Performance of \textsc{jd} modules for different network sizes of fixed density or fixed diameter \label{fig:shawn:jd}}
\end{center}
\end{figure}

The performance of the join decision modules is depicted in Fig.~\ref{fig:shawn:jd}. Again, the time efficiency of all modules is directly affected by the $k$-hop parameter (or $d$-hop parameter for \textsf{MaxMinD}). Similar results hold for the number of messages exchanged. As we can see, \textsf{MaxMinD} takes almost no time to complete and requires no messages. \textsc{dfs$\_$jd} needs more time than any other module as all available nodes have to be invited in a depth first manner and more interaction between the nodes take place. When using the fixed diameter topologies we can see that the increased node density has a great effect on the operation of \textsc{dfs$\_$jd} as clusters grow in size and more 
nodes have to be invited.

\begin{figure}
\begin{center}
\includegraphics[width=0.45\textwidth,keepaspectratio=true]{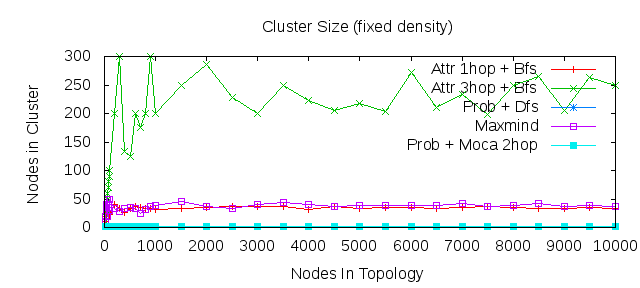}
\includegraphics[width=0.45\textwidth,keepaspectratio=true]{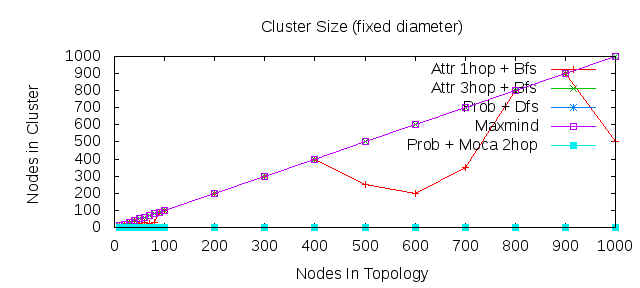}
\caption{Average size of the clusters created for different combinations of modules\label{fig:shawn:size}}
\end{center}
\end{figure}

Finally, we evaluate the size of the clusters formed using different combinations of modules. Fig.~\ref{fig:shawn:size} depicts the results for both types of topologies. As expected, in the fixed node density topologies all clustering schemes produce clusters with the same average cluster size. In the fixed diameter topologies, as the network diameter is small some algorithms 
like \textsf{MaxMinD} form a single cluster. As the network diameter increases, the number of clusters produced increases linearly.

In order to understand the performance of the modules in real hardware we continue our evaluation by conducting experiments in a local testbed comprised of up to $10$ iSense \cite{buschmann07isense} nodes placed on a $1$ hop network. These experiments allow us to measure real time and not simulated rounds. For each experiment, we assess the time required by each module to complete and the total number of messages exchanged. Compiling our application and using it with the iSense sensors required minimal modifications as everything was developed under the \textsf{Wiselib} library that fully supports the iSense platform.

\begin{figure}
\begin{center}
\includegraphics[width=0.45\textwidth,keepaspectratio=true]{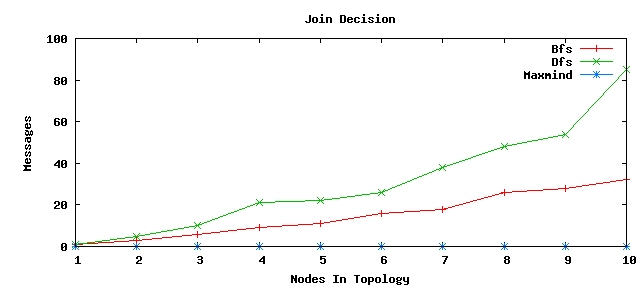}
\includegraphics[width=0.45\textwidth,keepaspectratio=true]{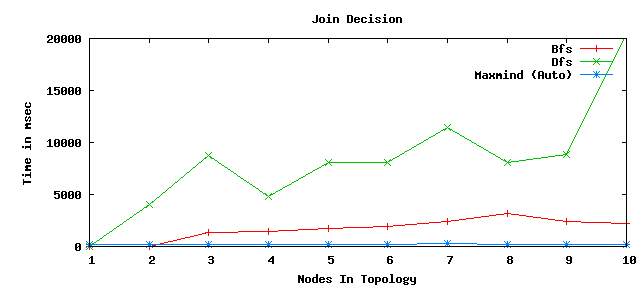}
\caption{\label{fig:isens:jd} Experimental evaluation of Time and Message efficiency of \textsc{jd} modules}
\end{center}
\end{figure}

In general, the results of all experiments were consistent with the results gathered during the simulator experiments. Of course, the data of the experiments is clearly in a smaller scale. Moreover, note that some minor differences in operation time were observed as perfect synchronization of the nodes and fully reliable communication channels are impossible to achieve in real environments. Due to space limitations we only include Fig.~\ref{fig:isens:jd} that depicts the results for the \textsc{jd} modules. Both \textsc{bfs} and \textsc{dfs} perform as expected and the second one needs more time to invite all $10$ nodes in the cluster. Clearly, the information gathering approach up to $d$-hops distance of \textsf{MaxMinD} hinders the overall performance of the system. Still, all three modules complete in very short time ($<20sec$) and short message exchanges ($<80$).

Apart from assessing the performance of the implemented modules, this study also demonstrates the benefit of our approach. With truly minimum effort (we just modify a parameter in the constructor of some modules and then recompile) we can examine the performance of a wide range of algorithms. Additionally, by mixing and matching base modules we can come up with new algorithms and easily compare their performance with previous versions. Essentially, this creates an ideal platform for comparing different ideas and design choices under a common framework as we can instantly test our ideas in simulated or real environments and easily fine tune the performance of the developed applications.

\section{Applications}

In the previous sections we discussed about the benefits of our approach in terms of reducing the implementation effort for developing clustering algorithms. We also show cased how to easily evaluate the performance of protocol variants or new protocols in a simulated and experimental framework. In this section, we discuss a third aspect of our approach that is of paramount importance to the development of WSN applications. 

Developing clustering algorithms by itself is of absolutely no use. The true purpose of organizing the network in clusters is to improve the scalability of the network when performing other operations such as data aggregation, routing, energy conservation etc.. Currently implemented algorithms are loosely coupled with routing protocols thus rendering them useless for improving the performance of other tasks. Our approach is totally different. The proposed architecture allows to include clustering algorithms as sub-components of other algorithms that deal with totally different problems. Interestingly, the clustering algorithms are tightly coupled with the higher-layer algorithm and in this way they can be easily exchanged with other ones. Essentially, this allows us to split the development and evaluation of our system in two different phases. We can first develop our system using one of the existing clustering algorithms and then during the evaluation, as demonstrated in the previous section, fine tune the clustering algorithm to maximize the performance of the higher level system.

\subsection{Hierarchical routing}

Clustering can improve the scalability of routing. After clusters are set-up, routing can be performed either in intra-cluster level or in inter-cluster level. In the first case, a node that needs to communicate with another one in the same cluster establishes a route to it (directly or through the cluster-head). In the second case, a node who wishes to reach another one that belongs to a different cluster, firstly communicates with its cluster-head who is now responsible to construct a route towards the destination's node cluster-head and then reach the final node through intra-cluster routing. Thus, one could observe that cluster-heads form an upper level of hierarchy that facilitates the routing process.

In order to address cluster-level routing in our component based approach, we propose the \textsc{ClusterRadio} component. This component consists of two mechanisms named \textsc{IntraClusterRadio} and \textsc{InterClusterRadio} that are built based on the information provided by the \textsc{it} component. \textsc{IntraClusterRadio} stores local information about nodes that belong to a certain cluster and it is able to set-up routes (if necessary) inside this cluster (e.g. route to the CH). The \textsc{InterClusterRadio} component is responsible for discovering a cluster's gateway nodes - nodes that are close and thus able to communicate with neighboring clusters. This way, when a node wants to reach a node at a different cluster, the cluster's gateway nodes forward the request to neighboring clusters searching for the desired cluster. After the route to the destined cluster has been setup the destination node can be reached through intra-cluster routing. 

The result of the above component is that the application developer can easily replace the flat routing algorithm with a cluster-based routing. Furthermore, the routing choice for inside the cluster and outside the cluster can be easily changed due to the component-based architecture.

\subsection{Group Key Establishment}

Another application of clustering is in the field of securing communication exchanges within the network by combining them with Group key establishment (GKE) algorithms. GKE is the procedure of setting up secret cryptographic keys between groups of nodes. This essentially guarantees to some extend the confidentiality and integrity of the information exchanged. A wide variety of GKE protocols based on asymmetric and symmetric cryptographic techniques has been proposed so far, e.g. \cite{DBLP:journals/scn/KotzanikolaouMVS09,DBLP:journals/ijahuc/LiYS10,DBLP:conf/comsware/DasS08}. 

The big challenges when designing a GKE protocol for sensor networks are scalability and efficiency \cite{DBLP:journals/entcs/ChatzigiannakisKLS07}. Although, the proposed protocols aim  at these goals, a GKE mechanism could execute faster and more efficiently when applied after a clustering algorithm has divided the network into clusters. Moreover, certain GKE protocols like \cite{1607595} are based on the assumption that the network is already organized into groups. Thus, one can realize that clustering techniques can improve the efficiency of computationally heavy protocols like GKE protocols and respectively the overall network performance. 

As an application example, we implement the GKE algorithm proposed in \cite{DBLP:journals/entcs/ChatzigiannakisKLS07}. It is based on Elliptic Curve Cryptograhy (ECC) and it employs a depth first traversal that visits all the network participants who contribute to the group key. To do this we register two \textit{callback} methods at the \textsc{CC} so that the common key is re-computed when each node joins the cluster and the key is finalized with the last node addition. The keys are generated using the ECC operations (point multiplication,encryption/decryption) provided by \textsf{Wiselib}. Then we use the \textsc{prob}, \textsc{dfs} and \textsc{norm} components to fix the operation of the cluster algorithm. The overall implementation requires a total of $11$ lines of code. We evaluate the performance of the resulting implementation in our iSense hardware. For the topology of $10$ nodes it requires approximately $7$ minutes to compute a common key of $163$ bits. Remark that ECC-based cryptography of 163-bit is equivalent to 1024-bit RSA keys.


In a very similar way, protocols like \cite{1607595,DBLP:journals/ijahuc/LiYS10} can be implemented by employing the necessary cryptographic mechanism and by using the information provided by the \textsc{it} component after the cluster formation has finished.

%

\section{Conclusions}

In this paper we study a large body of $90$ clustering algorithms as presented in $8$ recent surveys and tutorials. We focus on $23$ algorithms that are (i) widely studied by the relevant bibliography, (ii) appeared very recently in very relevant competitive conferences and/or (iii) are implemented in a well known WSN platform. We identify their main building blocks and present a very limited set of base modules that are parameterizable and can be interchanged to implement existing algorithms or new ones under study. This process is of great value for a generic algorithm engineering approach. 

We carefully select $5$ out of the $23$ algorithms and explain in details how we combine and parametrize the $6$ base modules to implement them. We conduct an extended simulation study to demonstrate the faithfulness of our implementations when compared to the original implementations. For all cases, our simulations are at very large scale thus also demonstrating the scalability of the original algorithms beyond their original presentations. Moreover, we conduct experiments and assess their practicality in real WSN. This also demonstrates the capability of our approach to make code easily available to the community that are hardware and OS independent.

Our modular architecture, the implementation of the algorithms using multiple components and the \textsf{Wiselib} environment provides a common platform so that comparisons between algorithms is easily accessible. Developers can easily mix and match modules in order to provide new clustering variants that best fit their application specifications. The ability to implement new algorithm variants with minimum effort is of significant importance for conducting experimental-driven research. 

We propose a component-based architecture for developing clustering algorithms that promotes exchangeability of algorithms and cross-layer implementations. We examine two important problems of hierarchical routing and group key establishment and show how they can be implemented by exploiting the proposed clustering architecture. We demonstrate how our approach makes the integration of algorithms more feasible. For the case of group-key establishment we also conduct experiments to assess the overall performance of the resulting system in a real environment.

Our study clearly demonstrates the applicability of our approach and the benefits it offers to both research \& development communities. Our code is open-source and is publicly available for download and use. Due to the blind-review process we omit the url of our code.

%
%

\section*{Acknowledgements}

This work has been partially supported by the European Union under 
contract numbers
ICT-258885 (SPITFIRE)

{\small
\bibliographystyle{abbrv}
\bibliography{main}
}

\end{document}